\documentclass[aps,prb,twocolumn]{revtex4}
\usepackage{graphics,epsfig,amsfonts,amssymb,amsmath}
\begin{document}

\title{Phenomenological description of the two energy scales in underdoped
superconducting cuprates}
\author{B. Valenzuela and E. Bascones}
\affiliation{Instituto de Ciencia de Materiales de Madrid,
CSIC. Cantoblanco. E-28049 Madrid, Spain}
\begin{abstract}
Raman and ARPES experiments have demonstrated that
in superconducting underdoped cuprates nodal and antinodal regions are
characterized by two energy scales instead of the one expected in
BCS. The nodal scale decreases with underdoping  while the antinodal one
increases. Contrary to the behavior expected for an
increasing energy scale, the antinodal Raman intensity decreases with
decreasing doping. Using the Yang, Rice and Zhang (YRZ) model, we show that
these features
%
%
are a consequence of the non-conventional nature of the superconducting state
in which superconductivity and pseudogap correlations are both present
and compete for the phase space.


\end{abstract}

\date{\today}
\pacs{74.72.-h,71.10.-w,78.20.Bh} \email{leni@icmm.csic.es,
belenv@icmm.csic.es}
\maketitle
The pseudogap  state of
underdoped  cuprates is characterized by a nodal-antinodal
dichotomy with Fermi arcs at the diagonals of the Brillouin zone
 (nodal region) and a gapped antinodal
region\cite{Timusk99,shen}.
Raman\cite{letacon06,devereauxrmp} and angle resolved photoemission
(ARPES)
experiments\cite{scgapmesot} have confirmed that a nodal-antinodal
dichotomy is also present in the superconducting
state\cite{panagopoulos}. These results suggest that the superconducting
state cannot be simply described by BCS theory, contrary to general
believe\cite{hTc}.

Inelastic Raman scattering probes the zero-momentum charge
excitations. The response of nodal ($\chi_{B_{2g}}$) and antinodal
regions ($\chi_{B_{1g}}$) can be separated. In the superconducting state
pair-breaking peaks appear in the spectra. As the normal state of
cuprates is characterized by a not yet understood broad continuum,
these peaks are better identified
in the
subtracted response in superconducting and normal
states
$\Delta\chi_{B_{1g,2g}}=\chi_{B_{1g,2g}}^{SC}-\chi_{B_{1g,2g}}^{N}$.
With a standard d-wave BCS gap $\Delta_S\cos(2\phi)$,
as generally used for cuprates,
the frequency and intensity
of these peaks in both $B_{2g}$ and $B_{1g}$
channels would be only controlled by $\Delta_S$.

The experiments\cite{letacon06} reveal that in underdoped cuprates
$\Delta\chi_{B_{1g}}$ and $\Delta\chi_{B_{2g}}$ show pair breaking
peaks with opposite evolution with doping, instead of the single
energy scale. The $B_{1g}$ peak shifts to higher energy and loses
intensity with underdoping, while the $B_{2g}$ peak shifts to lower
frequency without too much change in intensity. The Raman spectrum,
specially the intensity of the $B_{1g}$ peak has been one of the
experimental results more difficult to understand. A {\it modified}
BCS gap with higher harmonics\cite{letacon06}, and vertex
corrections\cite{vertex} or a very strongly anisotropic
renormalization of the quasiparticle\cite{letacon06} have been
invoked to explain the $B_{1g}$ peak frequency and intensity,
respectively.
\begin{figure}
\leavevmode
\includegraphics[clip,width=0.40\textwidth]{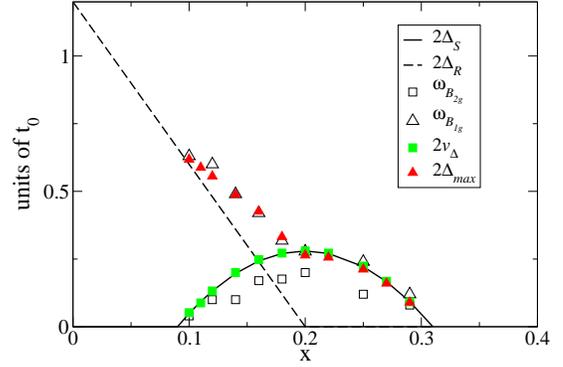}
\caption{(Color online) Comparison of pseudogap $\Delta_R$ and
superconducting $\Delta_S$ scales with ARPES nodal $v_\Delta$ and antinodal
$\Delta_{max}$ ones and with the frequency at which $B_{1g}$ ($\omega_{B_{1g}}$) and $B_{2g}$ ($\omega_{B_{2g}}$) Raman responses peak. The
parameters used depend on doping as, $\Delta_R(x)/2=0.3(1-x/0.2)$,
$\Delta_S(x)/2=0.07(1-82.6(x-0.2)^2)$, $t(x)=g_t(x)+0.169/(1+x)^2$,
$|t'(x)|=g_t(x)|t'_0|$ with $t'_0=-0.3$, $t''(x)=g_t(x)t''_0$ with
$t''_0=0.2$. All energies are in units of the bare hopping $t_0$. }
 \label{1}
\end{figure}

ARPES measurements\cite{scgapmesot} have also uncovered that
underdoping leads to an increase in the gap in the antinodal region
$\Delta_{max}$ and a decrease in the slope of the gap at the nodes,
$v_\Delta$, resulting in a U-shape dependence of the
gap\cite{scgapmesot} instead of the V-shape expected
from d-wave BCS. In the standard d-wave model, $\Delta_S$ gives both
$v_\Delta$ and $\Delta_{max}$. A single-energy scale  and a linear
(V-shape) dependence on $\cos(2 \phi)$ of the gap are observed in
the ARPES and Raman spectrum of overdoped
cuprates\cite{shen,devereauxrmp}.

In this letter, we show that the appearance of nodal and antinodal
energy scales and the suppression of intensity of $\Delta
\chi_{B_{1g}}$ with underdoping are a consequence of the
non-conventional nature of the superconducting (SC) state in which
superconductivity and pseudogap (PG) correlations are both present
and compete for the phase space and not just a modified BCS gap. We
assume that the PG strength, given by $\Delta_R$ vanishes at a
topological quantum critical point (QCP), and that the SC order
parameter $\Delta_S$ follows the critical temperature. Fig.~1 and
the evolution of the Raman intensity with doping $x$ are our main
results. Only the nodal energy scales follow the non-monotonic
dependence of the SC order parameter, being the slope of the gap at
the nodes $v_\Delta$  a good measure of $\Delta_S$. The antinodal
energy scale, i.e. the location of the $B_{1g}$ Raman peak
$\omega_{B_{1g}}$ and the maximum gap $\Delta_{max}$ in ARPES, is
intimately connected with the PG. The Raman spectra is calculated at
the bubble level. The decrease of intensity in the $B_{1g}$ channel
with underdoping is not due to vertex corrections, but to the
competition of PG and superconductivity in the antinodal region. The
renormalization of the quasiparticle weight $g_t$ just enhances this
effect.

We use the ansatz proposed by
Yang, Rice and Zhang\cite{YRZ} (YRZ) for the PG.
The YRZ model assumes that the PG can be described as a doped
spin liquid and proposes a phenomenological Green's function to
characterize it.
\begin{equation}
G^{YRZ}({\bf k},\omega)=\frac{g_t}{\omega-\xi_{\bf k}-\Sigma_R({\bf
k},\omega)}+G_{inc} \label{eq:green}.
\end{equation}
Here $\xi_{\bf k}=\epsilon_{0{\bf k}}-4t'(x)\cos k_x \cos
k_y-2t''(x)(\cos 2k_x+cos 2k_y)-\mu_p$, $\epsilon_{0{\bf
k}}=-2t(x)(\cos k_x+\cos k_y)$ and $\mu_p$ is determined from the
Luttinger sum rule. $\Sigma_R({\bf k},\omega)=\Delta_R({\bf
k})^2/(\omega+\epsilon_{0{\bf k}})$ diverges at zero frequency at
the umklapp surface $\epsilon_{0{\bf k}}$, and $\Delta_R({\bf
k})=\Delta_R(x)/2(\cos k_x -\cos k_y)$. The coherent part is similar
to the BCS diagonal Green's function with the non-trivial difference
that in BCS, the self-energy diverges at the Fermi surface (FS) and
not at the umklapp one. Besides there is no off-diagonal component
of the Green's function in our case and $\Delta_R$  does not break
any symmetry. For finite $\Delta_R$ the quasiparticle peak is split
into two and the FS consists of hole pockets close to $(\pm
\pi/2,\pm \pi/2)$. At $x_c$, $\Delta_R(x)$ vanishes at a topological
transition and a complete FS is recovered. Following  predictions of
mean field theory\cite{mft} the coherent spectral weight factor,
$g_t=2x/(1+x)$, decreases with underdoping and vanishes at half
filling. We use the same parameters proposed in the original
paper\cite{YRZ}, see Fig.~1.

Superconductivity is introduced in the standard way\cite{abrikosov}
as in Ref.\cite{YRZ}. The diagonal Green's function becomes
\begin{equation}
G^{RVB}_{SC}({\bf k},\omega)=\frac{g_t}{\omega-\xi_{\bf
k}-\Sigma_R({\bf k},\omega)-\Sigma_S({\bf k},\omega)}
\label{eq:greensc}.
\end{equation}
$\Sigma_S({\bf k},\omega)=|\Delta_S^2({\bf k})|/(\omega+\xi({\bf
k})+\Sigma_R({\bf k},-\omega)$) is the SC self energy with
$\Delta_S({\bf k})=\Delta_S(x)/2(\cos k_x -\cos k_y)$ with doping
dependence given in Fig.~1. Each quasiparticle
peak splits into four with energies $\pm E_{\pm}$.
\begin{eqnarray}
(E^\pm_{\bf k})^2&=&\Delta_{R{\bf k}}^2+\frac{\xi^2_{\bf
k}+\xi^2_{0{\bf k}}+\Delta_{S{\bf k}}^2}{2}\pm (E_{\bf k
}^{SC})^2\nonumber\\
(E_{\bf k}^{SC})^2&=&\sqrt{(\xi^2_{\bf k}-\xi^2_{0{\bf
k}}+\Delta_{S{\bf k}}^2)^2+4\Delta_{R{\bf k}}^2((\xi_{\bf
k}-\xi_{0{\bf k}})^2+
\Delta_{S{\bf k}}^2)} \nonumber
\end{eqnarray}
The spectral functions $A({\bf k},\omega)=-2{\rm Im} G({\bf
k},\omega)$ and $B({\bf k},\omega)=-2{\rm Im} F({\bf k},\omega)$
with $F({\bf k},\omega)$ the anomalous Green's function are
\begin{eqnarray}
 A({\bf k},\omega)&=&g_t \pi\{(v_{\bf k}^-)^2\delta(\omega+E_{\bf
 k}^-)+(u_{\bf k}^-)^2\delta(\omega-E_{\bf k}^-) {}\nonumber\\{}&&
 +(v_{\bf k}^+)^2\delta(\omega+E_{\bf k}^+)+(u_{\bf k}^+)^2\delta(\omega-E_{\bf
 k}^+)\},\nonumber\\
 B({\bf k},\omega)&=&g_t \pi\{u_{\bf k}^-v_{\bf k}^-(\delta(\omega+E_{\bf
 k}^-)+\delta(\omega-E_{\bf k}^-)) {}\nonumber\\{}&&
 +u_{\bf k}^+v_{\bf k}^+(\delta(\omega+E_{\bf k}^+)+\delta(\omega-E_{\bf
 k}^+))\}, \label{eq:akwsc}
\end{eqnarray}
with  coherence factors $ v_{\bf
k}^{2\pm}=\frac{1}{2}\Big( a^\pm_{\bf k}- b^\pm_{\bf k}/E^\pm_{\bf
k}\Big)$ and $ u_{\bf k}^{2\pm}=\frac{1}{2}\Big( a^\pm_{\bf k}+
b^\pm_{\bf k}/E^\pm_{\bf k})$, where $a^\pm_{\bf k}
=\frac{1}{2}(1\pm (\xi_{\bf k}^2-\xi_{\bf k0}^2+\Delta_{S{\bf k
}}^2)/E_{\bf k}^{SC})$ and $b^\pm_{\bf k} =\xi_{\bf k}a^\pm_{\bf k}
\pm \Delta_{R{\bf k}}^2 (\xi_{\bf k}-\xi_{\bf k 0})$.

In the bubble approximation\cite{mahan} the Raman response is
\begin{eqnarray}
\label{eq:raman} \lefteqn{{\rm
Im}\{\chi_{\gamma_\nu}(\Omega)\}=\sum_{\bf k} (\gamma_{\bf
k}^{\nu})^2 \int
\frac{d\omega}{4\pi}(n_F(\omega)-n_F(\omega+\Omega))
{}}\nonumber\\{}&& \{A({\bf k},\omega+\Omega)A({\bf
k},\omega)-B({\bf k},\omega+\Omega)B({\bf k},\omega)\}.
\end{eqnarray}
Here $n_F(\omega)$ is the Fermi function, $\nu=B_{1g},B_{2g}$ and
$\gamma^\nu_{\bf k}$, the $B_{1g}$ and $B_{2g}$ Raman vertices, are
proportional to $\cos k_x-\cos k_y$ and $\sin k_x \sin k_y$
respectively.

The  dependence on $x$ of the  nodal and antinodal subtracted Raman
spectra $\Delta \chi$ can be seen in Fig.~2a and 2b. Two different
energy scales $\omega_{B_{1g}}$ and $\omega_{B_{2g}}$ appear for
finite $\Delta_R$. These scales, signalled by an arrow in Fig.~2,
are plotted in Fig.~1. At $x_c=0.2$ the PG vanishes and
$\Delta\chi_{B_{1g}}$ and $\Delta\chi_{B_{2g}}$ peaks appear at an
energy close to $2 \Delta_S$. This is the expected behavior for a
BCS superconductor\cite{saddlepoint}. On the contrary, as $x$ is
reduced the $\Delta\chi_{B_{1g}}$ peak shifts to larger frequency
and its intensity decreases, while the $\Delta\chi_{B_{2g}}$ peak
shifts to lower frequency with a weakly $x$-dependent intensity.
This behavior resembles very much the experimental
one\cite{letacon06}. In Fig.~2a and 2b the spectra have been divided
by $g_t^2$ to emphasize that the suppression of the intensity in the
$B_{1g}$ channel with underdoping is not only due to the reduction
of the coherent part, as proposed in\cite{letacon06}. With $g_t^2$
included the weakening of the $B_{1g}$ transition with underdoping
is enhanced and the $B_{2g}$ signal decreases.
\begin{figure}
\leavevmode
\includegraphics[clip,width=0.40\textwidth]{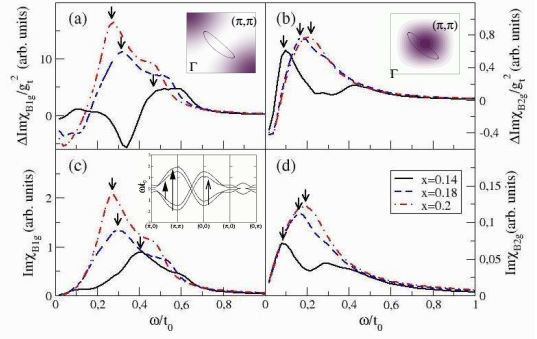}
\caption{ Insets in (a) and (b): Raman vertices $(\gamma_{\bf
k}^{B_{1g}})^2$ and $(\gamma_{\bf k}^{B_{2g}})^2$ in the first
quadrant of the Brillouin zone (BZ), with the Fermi pocket for $x=0.14$ drawn. Inset
in (c): bands for $x=0.14$ in the SC state. The arrows signal the
possible optical transitions. Main figures (a) and (b) subtracted
Raman response in $B_{1g}$ and $B_{2g}$ polarizations divided by the
coherent weight factor $g_t^2$, for $x=0.14$ and $x=0.18$  in the
underdoped regime and to $x_c=0.20$. (c) and (d) full Raman response
in the $B_{1g}$ and $B_{2g}$ channels respectively for the same
dopings. Raman units are the same for all dopings.
The arrows indicate the position, $\omega_{B_{1g}}$ and
$\omega_{B_{2g}}$, and intensity of the pair breaking peaks.
The $\delta$ functions in Eq.~\ref{eq:akwsc} have been
replaced by Lorenzians with a width of $0.05t(x)$ with $t(x)$ given
in Fig.~2a.
}
 \label{2}
\end{figure}

The appearance of two different energy scales for $B_{1g}$ and
$B_{2g}$ response is associated to the two pair-breaking transitions
in the inset of Fig.~2c with energies $2E_\pm(\bf{k})$. The
distinction between nodal and antinodal signal has its origin in the
coherence factors $u^2_\pm({\bf k})v^2_\pm(\bf{k})$ which weight
each transition. Shown in Fig.~3a (Fig.~3b) for $x=0.14$,
$u^2_-({\bf k})v^2_-(\bf{k})$ ($u^2_+({\bf k})v^2_+(\bf{k})$),
weights more heavily the nodal (antinodal) region. The $B_{2g}$ and
$B_{1g}$ spectra are respectively dominated by the transitions with
energy $2E_{-}$ and $2E_{+}$. The maxima in the Raman spectra in
Fig.~2a and 2b arise from the peaks in the densities of transitions
$N^{weight}_{2E_\pm}=\sum_{\bf k}u_\pm^2({\bf k})v_\pm^2({\bf k})
\delta(\omega - 2 E_\pm({\bf k}))$. In Fig.~3d $N^{weight}_{2E_-}$
peaks at a frequency smaller than $2\Delta_S$. $\omega_{B_{2g}}$
depends not only on $\Delta_S$ but also on the length of the Fermi
pockets along $(\pi,0)$-$(0,\pi)$ as its value comes from the edges
of the Fermi pockets. Due to the shrinking of the pockets with
increasing $\Delta_R$, $\omega_{B_{2g}}$ is shifted to lower
frequencies with underdoping, even if a doping independent
$\Delta_S$ is used. In Fig.~3e, the energy at which
$N^{weight}_{2E_+}$ peaks is determined by a saddle-point close to
$(\pi,0)$, along $(\pi,0)-(\pi,\pi)$ and is bigger than $2\Delta_S$.
In the UD region, this energy is mainly controlled by $\Delta_R$. If
$x \geq x_c$, only one of the pair-breaking transitions gives a
finite contribution for a given ${\bf k}$. A complete d-wave
superconducting gap is recovered and, as shown in Fig.~3c it is
tracked by the added coherence factors $u^2_+({\bf k})v^2_+({\bf
k})+ u^2_-({\bf k})v^2_-({\bf k})$.  The maximum in
$N^{weight}_{E_-}$ is not anymore a maximum of the total density
$N^{weight}_{tot}=N^{weight}_{E_-}+N^{weight}_{E_+}$, plotted in
Fig.~3f, but both densities of transitions match perfectly.
$N^{weight}_{ tot}$ is now the meaningful quantity. It peaks
close to $2\Delta_S$ recovering the BCS single-energy
behavior\cite{saddlepoint}.
\begin{figure}
\leavevmode
\includegraphics[clip,width=0.40\textwidth]{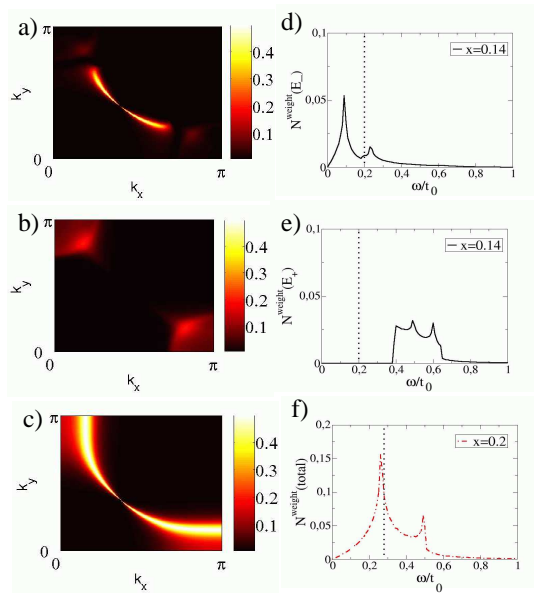}
\caption{
(a) to (c) coherence factors, in the first quadrant of the BZ, corresponding to
the pair breaking transitions with energies $2E_{\pm}$ (see text).
The weighted densities of
transitions $N^{weight}(E_\pm)$ are plotted in (d) to (f).
$2\Delta_S$ is marked with a dotted line.
The added coherence factors and total weighted density of
transitions for $x_c=0.2$ (critical doping) are shown in (c) and (e)
recovering the d-wave BCS result.} \label{3}
\end{figure}

A third {\it crossing} transition,  with energy
$E_-({\bf k})+E_+({\bf k})$, and larger intensity in the $B_{1g}$ channel is also allowed if $\Delta_R$ is finite,
as shown in the inset of Fig.~2c.
Its effect is small in  $\Delta\chi$, as
it is expected in both
the PG and the SC state.
The total responses
$\chi_{B_{1g}}^{SC}$ and $\chi_{B_{2g}}^{SC}$ including the
contribution of this crossing transition in the SC
state are plotted in Figs.~3c and 3d. It is not easy to distinguish
this transition from the pair-breaking ones.
To the best of our knowledge this transition has not been found yet in the
PG state  but it could be hidden in the broad background due
to strong inelastic scattering.

As expected for a gap with nodes, the $B_{2g}$ response in Fig.~3d is
linear at low frequencies. The slope is doping independent.
This independence, observed also
experimentally\cite{letacon06}, comes from a
cancellation between the dependencies of the quasiparticle weight
squared $g^2_t$, the SC order parameter, $\Delta_S$,
and the density of states, via the renormalization of the band
parameters.

The nodal and antinodal scales can be also seen in ARPES. As
discussed in ref.\cite{YRZ} for $x <x_c$ in the PG state, the FS
consists of closed hole pockets. However, due to the weak spectral
weight of the quasiparticle pole in the outer edge of the pocket,
the  ARPES spectra resembles the Fermi arcs observed experimentally.
The length of these arcs increases with doping, as seen in Fig.~4a
and Fig.~4b. A complete FS is recovered when $\Delta_R=0$ in
Fig.~4c. In the absence of a complete FS, to analyze the ${\bf
k}$-dependence of the gap we take the surface with maximal intensity
$\omega=0$ and $\Delta_S=0$. This surface, signalled in blue in the
pictures, resembles the one interpreted experimentally  as the
underlying FS. To compare with experiments\cite{scgapmesot} we
define $v_\Delta$ as the derivative of the energy with respect to
$\cos k_x -\cos k_y$ at the nodes\cite{vdelta} and $\Delta_{max}$ as
the maximum gap along this surface. Shown in Fig.~4d for $x=0.05$,
when $\Delta_S$ is zero but $\Delta_R$ finite, the energy vanishes
along the arc and a gap opens linearly with $\cos k_x-\cos k_y$ from
the arc edge. $\Delta_{max}$ increases with underdoping. A finite
$\Delta_S$ opens a gap along the arc in Fig.~4e. This gap depends
linearly on $\cos k_x-\cos k_y$, with slope $v_\Delta$ very close to
$\Delta_S$. Outside the arc\cite{intensity}, the gap depends on both
$\Delta_R$ and $\Delta_S$. In this UD SC region $v_\Delta$ increases
with doping and $\Delta_{max}$ decreases (see Fig.~1).
Correspondingly, the spectra does not depend linearly on $\cos k_x -
\cos k_y$ but has a U-shape dependence on $\cos k_x-\cos k_y$ with a
kink around the arc edge. Deviations from linearity increase with
underdoping. In Fig.~4f, at $x=x_c$, the linear V-shape BCS
dependence re-appears and $v_\Delta$ and $\Delta_{max}$ converge.
\begin{figure}
 \begin{center}
\leavevmode
\includegraphics[clip,width=0.40\textwidth]{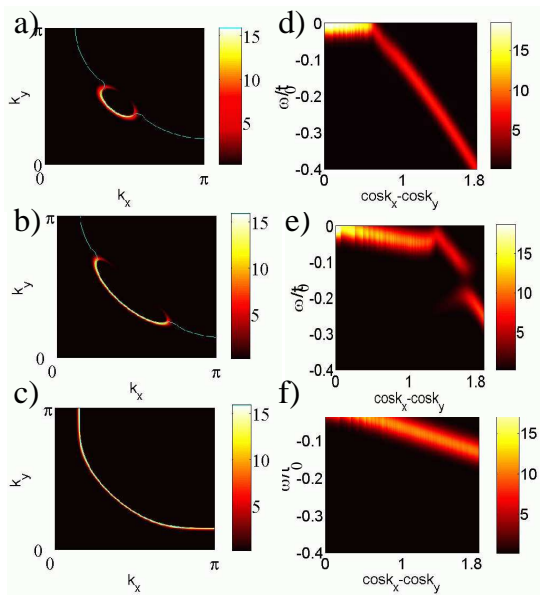}
\end{center}
\caption{(a) to (c) Map of  ARPES intensity in the first
quadrant of the BZ, at zero frequency and zero
$\Delta_S$ for (a) $x=0.05$, (b) $x=0.14$ and (c)
$x_c=0.20$. In blue, the maximum intensity surface.
(d) to (f) Energy spectrum corresponding to (d) x=0.05,
(e) x=0.14 and (f) x=0.20,  along the surface marked in (a) to (c). $\Delta_S$ is finite in (e)
and (f).
The $\delta$ functions in $A({\bf k},\omega)$,
have been replaced by lorenzians of width $0.001t_0$ and
the spectral
function  convoluted with a
gaussian of width $0.02t_0$ ($\sim 6-10 meV$) and a temperature $T=0.001t_0$.}
 \label{4}
\end{figure}

The evolution of ARPES scales $v_\Delta$ and $\Delta_{max}$, with
doping is plotted in Fig.~1, and compared with the ones found in
Raman and  the input $\Delta_S$ and $\Delta_R$.
The similarity with  experimental data is
striking\cite{letacon06}. With the $\Delta_S(x)$
used, $\omega_{B_{2g}}$ and $v_{\Delta}$ are non-monotonic on doping.
On the other
hand, the frequency at which $B_{1g}$ peaks
$\omega_{B_{1g}}$  follows very closely $2\Delta_{max}$ and both decrease  as $\Delta_R$ does.
Twice the gap
value at $(\pi,0)$, sometimes compared in the literature with
$\omega_{B_{1g}}$, is expected to be a bit larger than
$2\Delta_{max}$ plotted here. The nodal and antinodal scales merge
when the pseudogap correlations disappear.

In conclusion, we have reproduced the deviations from BCS in Raman
and ARPES experiments in  underdoped superconducting cuprates. Nodal
and antinodal energy scales
with opposite doping dependence
appear in both spectra. The nodal $B_{2g}$ response, peaks at a frequency
$\omega_{B_{2g}}$ which qualitatively follows the doping dependence
of the superconducting order parameter $\Delta_S$. On the contrary,
the energy of the pair breaking transitions in the antinodal region,
$\omega_{B_{1g}}$, decreases monotonically with increasing doping
and its intensity decreases with underdoping due to the competition
between pseudogap and superconducting correlations.
Twice the maximum value of the ARPES gap $\Delta_{max}$
along the maximum intensity surface follows very closely
$\omega_{B_{1g}}$. Within this model the slope of the gap at the
nodes, $v_{\Delta}$, as measured by ARPES, is a good measure of
$\Delta_S$ while the maximum value of the gap $\Delta_{max}$ arises
from an interplay between the pseudogap $\Delta_R$ and $\Delta_S$.
We emphasize that we have not tried to fit the experiments but just
taken the values proposed in the YRZ paper\cite{YRZ}.
For the values
used, $x_c$ and optimal doping coincide.
If this is not the
case\cite{loram}, it is $x_c$ which
controls the emergence of anomalous behavior.

Similar two-scale behavior could appear in other models with a QCP\cite{ddw}.
Possible differences could be the decreasing spectral weight with
underdoping, important for the constancy of the slope in
$B_{2g}$ channel.
which a priori is not expected in other QCP models. In the YRZ
ansatz the FS is truncated without breaking of symmetry and a
topological transition happens at $x_c$, in agreement with
experiments\cite{loram} and Dynamical Mean Field
Theory\cite{kotliar}. Our results suggest that the pseudogap is not
a precursor to the superconductivity but has a different origin and
persists in the superconducting state and that the antinodal scale
depends on both the pseudogap and the superconducting order
parameter. The smooth convergence of the antinodal scale with the
superconducting order parameter and a peak in $\Delta \chi_{B_{1g}}$
are hard to understand in models with separation in $\bf{k}$-space
in which antinodal quasiparticles, responsible of the pseudogap, do
not participate in superconductivity\cite{pines}. While not included
here, we believe that the appearance of two energy scales in the SC
state is robust enough to survive inelastic scattering.
\acknowledgments We thank M. Le Tacon, A. Sacuto, L. Tassini, R.
Hackl, J. Carbotte and G. Kotliar for  discussions and M.A.H
Vozmediano, A.V. Chubukov, F. Guinea and T.M. Rice for discussions
and reading of the preprint. Funding from MCyT  through
MAT2002-0495-C02-01, FIS2005-05478-C02-01 and Ramon y Cajal and from
Consejer\'ia de Educaci\'on de la Comunidad de Madrid and CSIC
through 200550M136 and I3P is acknowledged.



\begin{thebibliography}{99}
\bibitem{saddlepoint}
In the continuum the peak in $\Delta\chi_{B_{1g}}$ is at $2\Delta_S$. In
a lattice two peaks appear, the one at higher energy is due to band
structure effects and ignored in the discussion.
\bibitem{vdelta} $v_\Delta$ is
normalized to $\Delta_{max}$ for $\Delta(\phi)=\Delta_{max}
cos(2\phi)$\cite{scgapmesot}.
\end{thebibliography}
\end{document}